\documentclass[fullpage,11pt]{article}

\usepackage{color}
\usepackage{graphicx}
\usepackage{amsthm}
\usepackage{dsfont}
\usepackage{epsfig}

\newcommand{\old}[1]{{}}

\newtheorem{lemma}{Lemma}
\newtheorem{corollary}[lemma]{Corollary}
\newtheorem{theorem}[lemma]{Theorem}
\newtheorem{problem}[lemma]{Problem}

\def\paragraph#1{

{\bf #1}\hspace{.5em}}

\title{The Complexity of MaxMin Length Triangulation}

\author{
  S{\'a}ndor P.~Fekete\thanks{Algorithms Group, Department of Computer Science,
TU Braunschweig, 
38116 Braunschweig, Germany. E-mail: {\tt s.fekete@tu-bs.de} }}

\begin{document}

\maketitle

\begin{abstract}
In 1991, Edelsbrunner and Tan gave 
an $O(n^2)$ algorithm for finding the 
MinMax Length triangulation of a set of points in the plane.
In this paper we resolve one of 
the open problems stated in that paper, by
showing that finding a MaxMin Length triangulation
is an NP-complete problem. The proof implies that (unless P=NP),
there is no polynomial-time approximation algorithm that can approximate
the problem within any polynomial factor.
\end{abstract}

\section{Introduction}
Triangulating a set of points is one of the basic
problems of Computational Geometry: given a set $P$ of $n$ points in the
plane, connect them by a maximal set $\Delta$ of non-crossing 
line segments.  This implies that all bounded 
faces of the resulting planar arrangement are triangles, while the exterior
face is the complement of the convex hull of $P$. 

Triangulations are computed and used in a large variety of contexts, e.g.,
in mesh generation, but also as a stepping stone for other tasks. While it
is not hard to compute some triangulation, most of these tasks
require triangulations with special properties that should be optimized. 
Examples include maximizing the minimum angle, 
minimizing the total edge weight or the longest edge length.

In this paper we consider the task of computing a triangulation whose
shortest edge is as long as possible. We show that this problem
is NP-complete, resolving an open problem
stated by Edelsbrunner and Tan in 1991~\cite{et-qtaml-91}. The proof
implies that (unless P=NP), there cannot be any polynomial-time approximation
that gets with any polynomial factor.

\paragraph{Related Work.}
For a broad survey of triangulations in a variety of settings,
see the book~\cite{drs-t-10} by De Loeara, Rambau, and Santos.
Maximizing the minimum {\em angle} in a triangulation is achieved by
the Delaunay triangulation~\cite{d-sv-34}; making use of Fortune's sweepline
algorithm~\cite{f-savd-87}, it can be computed in
$O(n \log n)$. Minimizing the maximum {\em edge} can be computed in quadratic
time, as shown by Edelsbrunner and Tan~\cite{et-qtaml-91,et-qtaml-93}.
One of the most notorious problems regarding triangulations was finally
resolved by Mulzer and Rote~\cite{mr-mwtnh-08}, who proved that
finding a triangulations of minimum total edge length (a {\em minimum-weight
triangulation}) is an NP-hard problem. As shown by Remy and
Steger~\cite{rs-qptas-09}, there is a PTAS for this problem.
The maximum-weight triangulation problem has been considered
by Qian, and Wang~\cite{qw-ltasm-04}, who gave a linear-time approximation
scheme for the case of a ploint set in convex position, and
by Chin, Qian, and Wang~\cite{cqw-pmwt-04}, who gave a 4.238-approximation
algorithm. Computing a MaxMin triangulation has been considered
by Hu~\cite{h-ltaml-07}, who gave a linear-time algorithm for a
convex polygon (and thus for a sorted set of points in convex position),
and proved that the graph version of the problem is NP-hard.
Schmidt~\cite{s-mltp-12} showed that finding a geometric
MaxMin triangulation is NP-complete in the presence of obstacles,
such as inside of a polygon with holes; she also showed that computing a
MaxMin triangulation for a simple polygon can be solved in polynomial time
by making use of dynamic programming.

\section{NP-Completeness}
\subsection{An Auxiliary Problem}

We start by showing that the following auxiliary problem is NP-complete, based
on a reduction of Planar 3Sat. 

\begin{problem}
{\sc Covering by Disjoint Segments} (CDS)

{\bf Given:} A specified set $S$ of line segments
(``stabbers'') in the Euclidean plane,
and a subset $T$ of their intersection points (``targets''). 

{\bf Wanted:}
A non-intersecting subset of the stabbers that covers all targets.
\end{problem}

This problem is somewhat related to one considered by Megiddo and Tamir~\cite{mt-cllfp-82}, who showed that it is NP-hard to compute the minimum number
of straight lines that are necessary to cover a given set of points in the
plane. Note, however, that CDS considers
line segments of finite length that are required to be disjoint.

\begin{lemma}
\label{le:aux}
The problem CDS is NP-complete.
\end{lemma}

\begin{figure}[h]
\centering
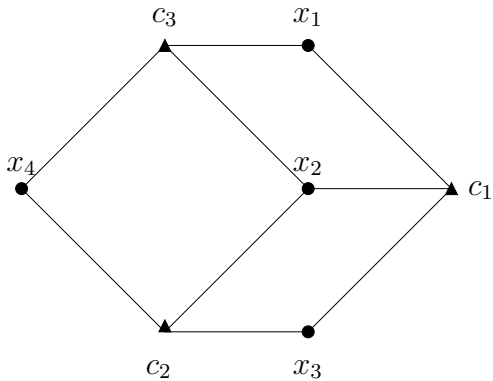
  \caption{A planar straight-line drawing of the variable-clause incidence
graph $G_I$ of the {\sc Planar 3SAT} instance
$I=(x_1\vee x_2\vee \overline{x_3})
\wedge(\overline{x_2}\vee x_3\vee\overline{x_4})
\wedge(\overline{x_1}\vee x_2\vee x_4)$. }
\label{fig:3sat}
\end{figure}

\proof
We give a reduction of {\sc Planar 3SAT}, which is the subclass of
3-satisfiability problems for which the variable-clause incidence graph
is planar---see Figure~\ref{fig:3sat}. To this end, we start
with an arbitrary 3SAT instance $I$ in conjunctive normal form, and
use it to construct a CDS instance $(S,T)_I$ that is solvable
if and only if $I$ can be satisfied.

\begin{figure}[h]
\centering
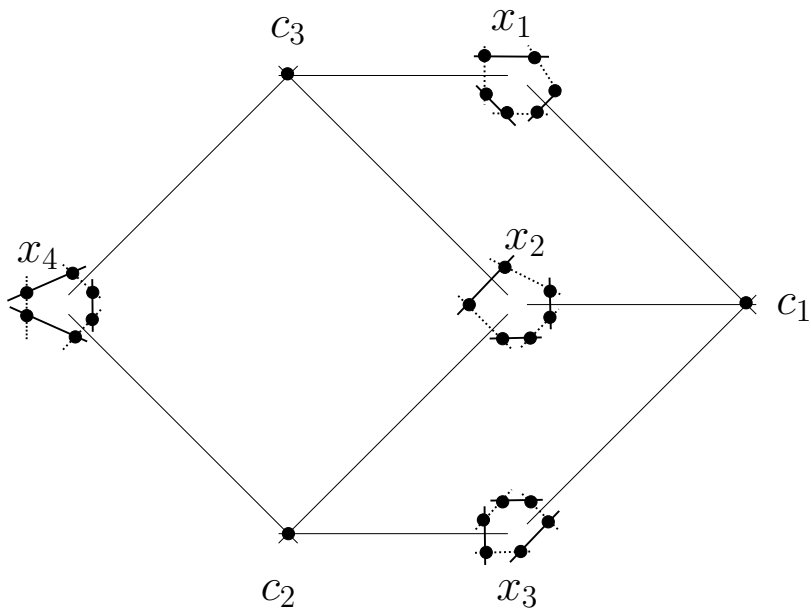
  \caption{A CDS instance of the 3SAT instance $I$. Even variable segments 
are shown in bold, odd variable segments are dotted; clause segments are shown
in thin solid. Dots indicate the target points. Labels indicate the original
locations of the variable and clause vertices in the drawing of $G_I$.
}
\label{fig:cds}
\end{figure}

In the following, we describe details of the construction; an example
of the resulting arrangement is shown in Figure~\ref{fig:cds}.

As a first step, the variable-clause incidence graph $G_I$ is embedded
into the plane, such that all vertices have integer
coordinates bounded by $O(n)$ and the resulting edges are represented
by line segments;
this can be achieved
in polynomial time by a variety of graph-drawing algorithms.
Let $V_{\mbox{var}}$ be the set of vertices that represent variables,
let $V_{\mbox{cla}}$ be the set of vertices that represent clauses,
and let $E_{\mbox{cla}}$ be the 
set of line segments that represent edges in $G_I$.
Scaling the resulting graph layout by a factor of $O(n^2)$ 
results in an arrangement in which each pair of vertices are at
least a distance of $\Omega(n^2)$ apart. 

In a second step, we replace each variable vertex $v_x\in V_{\mbox{var}}$ 
by an even cycle
of $O(n)$ intersecting ``variable''
line segments surrounding the original vertex 
at distance $\Theta(n)$. Each cycle consists of $\delta(v)$ ``even'' and
$\delta(v)$ ``odd'' variable segments;
their precise location and parity 
is chosen in a way that ensures that an edge $\overline{v_x v_c}$ in $G_I$ intersects an
odd segment iff variable $x$ occurs in clause $c$ in an unnegated fashion,
and an edge $\overline{v_x v_c}$ in $G_I$ intersects an
even segment iff variable $x$ occurs in clause $c$ in a negated fashion.
Let $S_{\mbox{var}}$ be the resulting set of variable segments.
Moreover, let $S_{\mbox{cla}}$ be the set of line segments obtained by
shifting all segments in $E_{\mbox{cla}}$ by a distance of $\Theta(n)$ towards 
its clause endpoint,
such that intersection with an appropriate variable
segment is maintained. Let $S=S_{\mbox{var}}\cup S_{\mbox{cla}}$.

Now let $T_{\mbox{var}}$ be the set of intersection points of variable
segments, let $T_{\mbox{cla}}$ be the set of intersection points of clause
segments, and  $T=T_{\mbox{var}}\cup T_{\mbox{cla}}$.
We claim: There is a subset $C\subset S$ that covers all points in $T$,
if and only if there is a satisfying truth assignment for $I$.

For the ``if'' part, consider the truth assignment of a variable $x$.
If $x$ is set to be true, choose all the odd variable segments for $x$;
if $x$ is set to be false, choose all the even variable segments for $x$.
In either case, all intersection points of the variable segments for $x$
are covered. Because every clause $c$ must have a satisfying literal,
picking the segment that connects the clause vertex with the corresponding variable does not intersect one of the selected variable segments.

For the converse ``only if'' part, consider a set $C\subset S$
of non-crossing segments that covers all points in $T$.
First it is easy to see by induction that if for some $x$, $C$ contains any
even segments, it must contain all even segments; otherwise, it must contain all
odd segments. This induces a truth assignment for all variables. 
Now it is easy to see that a clause vertex can only be covered by a clause
segment that does not cross a variable segment, which implies a satisfying
truth assignment.
\qed

\subsection{Hardness of MaxMin Triangulations}
Before exploiting the construction of Lemma~\ref{le:aux} for the main result,
we note a helpful lemma; the proof is elementary.

\begin{lemma}
\label{le:sep}
Let $P$ be a set of points in the plane, and let $p_i, p_j\in P$.
A triangulation $\Delta$ contains the edge $\overline{p_i p_j}$, iff
there is no edge in $\Delta$ that separates $p_i$ from $p_j$.
\end{lemma}

Now we proceed to the main theorem.

\begin{theorem}
\label{th:main}
It is NP-hard to decide whether a set $P$ of $n$ points in 
the plane has a triangulation
with smallest edge of length at least $c$, for some positive number $c$.
\end{theorem}

\proof
Consider the arrangment constructed for the proof of Lemma~\ref{le:aux}.
Let $Q$ be the set of all end points of segments in $S$, 
and $T$ be the set of target points.
We perturb all points in $Q\cup T$ 
by appropriate powers of $1/n$, such 
that the only triples of collinear points correspond to segments
in $S$ with their covered points.
For simplicity, we continue to 
refer to the resulting sets as $T$ and $S$. As a result, we get
a set of points and line segments, such that any triangle formed by
three points or a segment and a point not on the segment has smallest
height at least some $\delta>0$.

\begin{figure}[h]
\centering
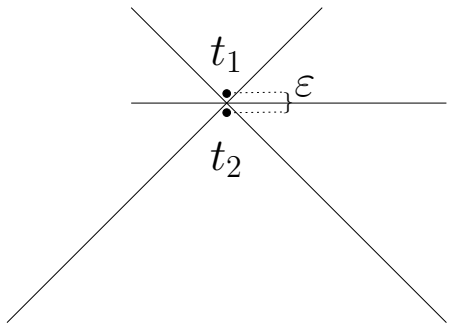
  \caption{Replacing the target points of the CDS instance by point pairs at distance $\varepsilon$.}
\label{fig:eps}
\end{figure}

Furthermore,
the segments covering a target point $t\in T$ subdivide its neighborhood
into four (for $t$ on a variable cycle) or six (for $t$ at a clause vertex)
sectors. Replace each point $t\in T$ by a pair of points $t_1, t_2$
at an appropriately
small distance $\varepsilon<<\delta$ in opposite sectors. 
See Figure~\ref{fig:eps}.

We claim: there is a triangulation with shortest edge of length greater than
$\varepsilon$, iff the corresponding CDS instance can be solved.

For the ``if'' part, note that by construction,
any segment that covers a target point $t$ must separate the corresponding 
point pair $\{t_1, t_2\}$. As all points in $T$ are covered, all 
close pairs are separated, and the claim follows from Lemma~\ref{le:sep}.

Conversely, any pair $\{t_1, t_2\}$ at distance $\varepsilon$ corresponding to
a target point $t$ must be separated.  By construction (and the perturbation
argument), a line segment $\ell$ that connects two points $p_1$ and $p_2$ can
only get within distance $\delta$ of $t$ if $\ell$ covers $t$ in the CDS
construction. This induces a solution to the CDS instance.
\qed

We note an important implication of our construction.

\begin{corollary}
\label{co:inapprox}
Let $p(x)$ be some polynomial. Then the existence of a polyno\-mi\-al-time
algorithm that yields a $p(n)$-approximation for the problem of finding
a MaxMin length triangulation implies P=NP.
\end{corollary}

\proof
In the proof of Theorem~\ref{th:main}, choose $\varepsilon$ small enough
that $\delta/\varepsilon>p(n)$. Then a $p(n)$-approximation requires
finding a triangulation in which all the $\varepsilon$-edges are intersected.
\qed

\section{Conclusions}
Even though our proof implies that finding an approximately
maxmin triangulation in deterministically polynomial time is a hopeless
task, there are a number of interesting issues that remain. These
include practically useful methods for constructing
exact or approximately optimal solution, as well as positive results
for special cases and variations.

\section*{Acknowledment}
Thank you to Christiane Schmidt and Joe Mitchell for helpful conversations.

\bibliographystyle{abbrv}
\bibliography{lit}

\end{document}